\documentclass[conference]{IEEEtran}
\IEEEoverridecommandlockouts
\usepackage{cite}
\usepackage{amsmath,amssymb,amsfonts}
\usepackage{algorithmic}
\usepackage{graphicx}
\usepackage{textcomp}
\usepackage{xcolor}
\usepackage{comment}
\usepackage{makecell}
\usepackage{multirow}
\usepackage{xurl}
\usepackage{float}
\usepackage{fancyhdr}
\usepackage{hyperref}
\usepackage{booktabs}           
\hypersetup{
  colorlinks, linkcolor=blue
}
\usepackage{longtable}
\def\BibTeX{{\rm B\kern-.05em{\sc i\kern-.025em b}\kern-.08em
    T\kern-.1667em\lower.7ex\hbox{E}\kern-.125emX}}
    
\renewcommand{\baselinestretch}{1} 

\makeatletter
\newcommand*\titleheader[1]{\gdef\@titleheader{#1}}

\AtBeginDocument{%
  \let\st@red@title\@title%
  \def\@title{%
    \bgroup\normalfont\large\centering\@titleheader\par\egroup
    \vskip1.5em\st@red@title}
}
\makeatother

\title{Arrhythmia Classification Using CGAN-Augmented ECG Signals*\\

\thanks{*This work was partially supported by Open Cloud Institute (OCI) at UTSA and the National Science Foundation under Grant No. CNS-2125643.}
}

\titleheader{This paper has been presented and published in IEEE BIBM Conference 2022 (Las Vegas)}

\begin{document}




\author{\IEEEauthorblockN{Edmond Adib}
\IEEEauthorblockA{\textit{\footnotesize Electrical and Computer Engineering Department} \\
\textit{\small University of Texas at San Antonio}\\
\small San Antonio, USA \\
edmond.adib@utsa.edu}

\and

\IEEEauthorblockN{Fatemeh Afghah}
\IEEEauthorblockA{\textit{\footnotesize Electrical and Computer Engineering Department} \\
\textit{\small Clemson University}\\
\small Clemson,USA \\
fafghah@clemson.edu}

\and

\IEEEauthorblockN{John J. Prevost}
\IEEEauthorblockA{\textit{\footnotesize Electrical and Computer Engineering Department} \\
\textit{\small University of Texas at San Antonio}\\
\small San Antonio, USA \\
jeff.prevost@utsa.edu}

}

\maketitle

\begin{abstract}
ECG databases are usually highly imbalanced due to the abundance of Normal ECG and scarcity of abnormal cases. As such, deep learning classifiers trained on imbalanced datasets usually perform poorly, especially on minor classes. One solution is to generate realistic synthetic ECG signals using Generative Adversarial Networks (GAN) to augment imbalanced datasets. In this study, we combined conditional GAN with WGAN-GP and developed AC-WGAN-GP \emph{in 1D} form for the first time to be applied on MIT-BIH Arrhythmia dataset. We investigated the impact of data augmentation on arrhythmia classification.  Two models were employed for ECG generation: \emph{(i)} unconditional GAN; Wasserstein GAN with gradient penalty (WGAN-GP) is trained on each class individually; \emph{(ii)} conditional GAN; one Auxiliary Classifier WGAN-GP (AC-WGAN-GP) model is trained on all classes and then used to generate synthetic beats in all classes. Two scenarios are defined for each case: \emph{(a)} unscreened; all the generated synthetic beats were used, and \emph{(b)} screened; only high-quality beats are selected and used, based on their Dynamic Time Warping (DTW) distance to a designated template. The state-of-the-art ResNet classifier (\textit{EcgResNet34}) is trained on each of the four augmented datasets and the performance metrics (precision/recall/F1-Score micro- and macro-averaged, confusion matrices, multiclass precision-recall curves) were compared with those of the original imbalanced case. We also used a simple metric Net Improvement. All the three metrics show consistently that unconditional GAN with raw generated data creates the best improvements.
\end{abstract}


\section{Introduction}
One of the major causes of death is cardiovascular diseases. In 2019, it reached 32\% of all deaths worldwide\cite{6184767}. ECG is widely used in the diagnosis of cardiovascular diseases mostly since it is  non-invasive and painless. Diagnosis is usually performed by human specialists which is time-consuming and prone to human error, in case of availability. However, automatic ECG diagnosis is becoming increasingly more acceptable since not only it eliminates randomized human errors, but also it can be available as a bedside testing any time and anywhere using common and affordable wearable heart monitoring devices. Automatic ECG diagnosis algorithms are usually deep neural network (DNN) classifier models which classify the ECG beats depending on the general pattern of the ECG heartbeat. Datasets used for training these classifiers are usually highly imbalanced, due to the fact that normal beats are more abundant and some of the abnormal patterns are so scarce that the classifier can hardly train on them. Moreover, the ECG signals associated with different arrhythmia can vary significantly across different patients, meaning that some patterns are rarely seen in training datasets. Thus, the trained classifier often performs poorly, especially on minor classes. 

There are several approaches to address the imbalance in datasets. One approach is data augmentation by synthetic data generated by deep generative models. Generative Adversarial Networks (GAN) \cite{goodfellow2014generative} and Variational Auto-Encoder (V/AE) \cite{doersch2016tutorial} models have been used to generate synthetic data \cite{frid2018synthetic}, \cite{jordon2018pate}. Adib et al. \cite{adib2021synthetic} compared the performance of several GAN models in the generation of synthetic realistic ECG signals. Another option is to use oversampling methods such as SMOTE \cite{chawla2002smote} to generate synthetic data. Other approaches to deal with inadequacy of samples are to utilize new designs of loss functions such as focal loss \cite{lin2017focal} or using a new training scheme, e.g., few-shot training \cite{wang2020generalizing}.

In this study, we developed a 1D conditional GAN (AC-WGAN-GP) and applied it to generate ECG signals (single heartbeat time series) for the first time. We investigated the impact of data augmentation on the performance of classification of imbalanced datasets. The MIT-BIH Arrhythmia dataset is used to train two models from the GAN family: \textbf{\emph{(a)}} \textit{unconditional GAN} (Wasserstein GAN with gradient penalty, WGAN-GP) and \textbf{\emph{(b)}} the developed \textit{conditional GAN} (Auxiliary Classifier Wasserstein GAN with gradient penalty, AC-WGAN-GP). In each case, the generated data are used in two scenarios: \textbf{\emph{(i)}} \emph{unscreened} or raw, i.e., without any further processing and \textbf{\emph{(ii)}} \emph{screened}, i.e., only a subset of generated beats is used: each generated beat was measured up against the corresponding template and discarded if its distance was above a certain threshold. Then, the state-of-the-art classifier (\textit{EcgResNet34}) was used to investigate the impact of augmentation on the classification metrics and confusion matrix. To compare the cases, micro-averaged Precision score, multi-class Precision-Recall curves and the \emph{net improvement} in True Positives (confusion matrix' main diagonals) are used. Generating synthetic ECG signals provides a richer data set which is potentially expandable indefinitely, while the  other options (new design of loss functions or new scheme of training) only soften the negative impact of imbalance in datasets. To the best of our knowledge, this is the first time that a \emph{1D} AC-WGAN-GP model is developed and applied on the benchmark MIT-BIH dataset with the intention of enrichment of the dataset.

\subsection{Related Works}
Wang et al. \cite{Wang2019ECGAD} used the synthetic beats generated by a 14-layer ACGAN to augment their dataset. Delaney et al. \cite{delaney1909synthesis} studied the generation of realistic synthetic signals using a range of architectures from the GAN family. They used \emph{Maximum Mean Discrepancy} (MMD) and \emph{Dynamic Time Warping} (DTW) as metrics to quantitatively evaluate the generated beats. Hyland et al. \cite{esteban2017real} employed the same architecture (two-layer BiLSTM) in the generator and the discriminator to generate synthetic ECG signals. They used MMD and two other innovative metrics  to evaluate the quality of the generated beats. Wulan et al. \cite{wulan2020generating} used a multiclass DCGAN model to generate 3 classes of ECG. Zhu et al. \cite{zhu2019electrocardiogram} devised a novel BiLSTM-CNN GAN to generate synthetic ECG signals. Zhang et al. \cite{zhang2021synthesis} used the standard 12 lead ECG signals on a novel GAN model. 
None of the above works implemented WGAN-GP or AC-WGAN-GP, only  \cite{Wang2019ECGAD} used AC-GAN model. 

\section{Proposed Method}

\subsection{Background}


Generative models find the distribution of the original data either implicitly (e.g., in GAN models \cite{goodfellow2014generative}) or explicitly (e.g., in V/AE models \cite{doersch2016tutorial}).

\subsubsection{\textbf{Generative Adversarial Networks}}
In general, GAN architectures are comprised of two blocks of networks: the generator and the discriminator \cite{goodfellow2014generative}. These two blocks compete with each other in a two-player zero-sum game with a loss function of $V(G, D)$ (\ref{eq:GAN ValF}), in which the generator tries to generate fake data so close to the original real data distribution that the discriminator cannot distinguish from the real data. The discriminator is a binary classifier that is trained on real and fake data to classify the inputs as real or fake. 
\vspace{-3mm}
\begin{multline}
	\label{eq:GAN ValF}
 	\min_{G} \max_{D} V(G, D) =  
 	E_{x\sim P_{data} (X)}[log D(X)] \\ +  E_{z\sim P_{z}(z)}[log(1-D(G(z)))]
\end{multline}

The following types of GAN family are utilized in this study as described below:

\paragraph{\textbf{WGAN-GP}}
It can be shown that $V(G, D)$ in (\ref{eq:GAN ValF}) reduces to Jensen-Shannon distance between the distributions of the real and fake data when the discriminator is at its optimum \cite{arjovsky2017wasserstein}. Jensen-Shannon distance function is not a smooth function and is not differentiable at zero, whereas the Wasserstein distance function has a smooth behavior and is differentiable everywhere. Moreover, it prevents the mode collapse which is a very common issue in regular GAN. All these lead to a better and more stable training. 

The necessary condition of \emph{k}-Lipschitz continuity is satisfied in WGAN by a technique called Parameter Clipping \cite{arjovsky2017wasserstein}, i.e., keeping the magnitude of parameters of the model bounded, which can easily lead to vanishing/exploding gradient. In WGAN with gradient penalty (WGAN-GP) \cite{gulrajani2017improved}, the \emph{k}-Lipschitz continuity constraint is satisfied by regularizing the loss function to keeping the norm of the gradient below 1.

\paragraph{\textbf{AC-WGAN-GP}}
WGAN-GP is an unconditional model i.e., the probability distributions used in the loss function are not conditional. Thus, all the data (training and generated) are in one class. However, if the probabilities in the loss function in (\ref{eq:GAN ValF}) are conditioned on the labels, the model becomes multiclass \cite{gulrajani2017improved}.
\begin{multline}
	\label{eq:CGAN ValF}
 	\min_{G} \max_{D} V(G, D) = 
 	E_{x\sim P_{data} (X)}[log D(X|y)] \\ +  E_{z\sim P_{z}(z)}[log(1-D(G(z|y)))]
\end{multline}

\subsubsection{\textbf{Classifier, {{ECG}}ResNet34}}
ResNet34 \cite{he2016deep} is a 34-layer model and is the state-of-the-art in classification of images (\emph{2D}). It incorporates residual building blocks following the residual stream logic: $F(x)+x$. Each building block is comprised of two $3\times3$ convolutional layers where the residual stream, $x$, goes directly from the input to the outlet of the block that prevents deterioration of training accuracy in deeper models \cite{he2016deep}. This classifier is pretrained on the ImageNet dataset (more than $100,000$ images in $200$ classes). We used its \emph{1D} implementation (\hspace{1sp}\cite{layshukecgresnet34}) to classify the heartbeats.

\subsection{Dataset and Segmentation}

One of the most used benchmark datasets in ECG signal analysis is MIT-BIH Arrhythmia \cite{moody2001impact} \cite{goldberger2000components}. It contains $48$ recordings, each $30$ minutes long with two channels. The analog signals are digitized at $360$ Hz. In this study, the long ECG signals are segmented to individual heartbeats using an adaptive window method which takes into account the changes in heartbeat rates. The R-peak of the QRS complex of heartbeats are annotated. For each beat, the distances to the next and previous R-peaks are determined using the annotation and $75\%$ of each is used as the cutoff to find the boundaries of the beat. The resulted dataset is highly imbalanced which is comprised of $109,338$ individual beats in $15$ classes. For this study, all the beats are resampled to $256$ timesteps per beat.

\subsection{Model Designs}

Two models have been used to generate beats: \textit{unconditional} GAN (WGAN-GP) and the developed \textit{conditional} GAN (AC-WGAN-GP). WGAN is chosen because of its better behavior in training and mode collapsing prevention.

\paragraph{Unconditional GAN}

The architectures of the generator and the critic are comprised of building blocks, which are repeated multiple times. In the generator they are comprised of: \textbf{\emph{(1)}} 1D transpose convolution layer (ConvTranspose1d) with a kernel size of $4$, a stride of $2$ and padding of $1$, \textbf{\emph{(2)}} a $1$D batch normalization (BatchNorm1d) and \textbf{\emph{(3)}} a rectified linear unit (ReLU). In the critic they are  comprised of: \textbf{\emph{(1)}} a $1$D convolutional layer with a kernel size of $4$, a stride of $2$ and padding of $1$, \textbf{\emph{(2)}} a $1$D instance normalization layer (InstanceNorm1d), and \textbf{\emph{(3)}} a Leaky ReLU (LeakyReLU). The details of the architectures are shown in Table \ref{tab:archs_wgangp}. The numbers in the brackets are the dimensions of the output from the layer. 


\begin{table}[ht]

	\scriptsize
 	\caption{\textit{\textbf{Unconditional GAN Architecture}}}

 	\centering
 		\begin{tabular}{| c | c | c | }
 		\hline
 			\thead{\textbf{Layer}} & \thead{\textbf{Generator}} & \thead{\textbf{Critic}} \\
 			\hline
 			\hline
 				\makecell{Input} & \makecell[cc]{$16\times100\times1$}  & \makecell[cc]{$16\times1\times256$}\\
 			    \hline
 			
     			\makecell{1} & \makecell[cc]{Block}  & \makecell[cc]{Vonv1d, LeakyReLU}\\
     			\hline
     			
     			\makecell{2} & \makecell[cc]{Block}  & \makecell[cc]{Block}\\
     			\hline
     			
     			\makecell{3} & \makecell[cc]{Block}  & \makecell[cc]{Block}\\
     			\hline
     			
     			\makecell{4} & \makecell[cc]{Block}  & \makecell[cc]{Block}\\
     			\hline
     			
     			\makecell{5} &  \makecell[cc]{ConvTranspose1d} & \makecell[cc]{Conv1d}\\
     			\hline 
     			
     			\makecell{6} &  \makecell[cc]{FC (64, 256)} & \makecell[cc]{FC (64, 256)}\\
     			\hline 
     			
     			\makecell{7} &  \makecell[cc]{tanh} &  - \\ 
     			\hline
     			
     			\makecell{Output} &  \makecell[cc]{$16\times1\times256$} &  \makecell[cc]{$16\times1\times1$} \\ 
     			\hline 
 			
 		\end{tabular}		
 	
 	\label{tab:archs_wgangp}
 
\end{table}

\paragraph{Conditional GAN}

We combined 1D conditional GAN with WGAN-GP to develop \textbf{\emph{1D}} AC-WGAN-GP. Class labels are embedded to a dimension of $(100\times1)$ and concatenated with the latent variable with the same dimension before feeding to the generator. The concatenated input to the generator has a dimension of $(200\times1)$. In the critic, the labels are embedded to a dimension of $(1\times256)$ and then concatenated to the heartbeat signal (fake or real) with the dimension of $(1\times256)$ \emph{as a channel} to produce an input of $(2\times256)$ befor being fed to the critic. The same building blocks in the generator and the critic of WGAN-GP are used correspondingly in this model too. The generator and critic parameters are initialized from a normal distribution with zero mean and a standard deviation of $0.02$. The Adam optimizer was used with a learning rate of $0.0001$. The details of the rest of the architectures are shown in Tables \ref{tab:archs_acwgangp_gen} and \ref{tab:archs_acwgangp_critic}. 

\begin{table}[ht]
	\scriptsize
 	\caption{\textit{\textbf{Conditional GAN Architecture - Generator}}}
 	\centering
 		\begin{tabular}{| c | c | c | }
 		\hline
 			\thead{\textbf{Layer}} & \multicolumn{2}{c|}{\thead{\textbf{Generator}}} \\
 			\hline
 			\hline
 				\makecell{Input} & \multirow{3}{*}{$16\times100\times1$ (feature)}  & \makecell[cc]{16 (label)}\\
 				\cline{1-1}\cline{3-3}

     			\makecell{1} & & \makecell[cc]{embedding ($16\times100$)}\\
     			\cline{1-1}\cline{3-3}
     			
     			\makecell{2} &   & \makecell[cc]{reshape ($16\times100\times1$)}\\
     			\hline
     			
     			\makecell{3} & \multicolumn{2}{c|}{Concatenate ($16\times200\times1$)} \\
     			\hline
     			
     			\makecell{4} & \multicolumn{2}{c|}{block ($16\times1024\times1$)} \\
     			\hline
     			
     			\makecell{5} & \multicolumn{2}{c|}{block ($16\times512\times8$)} \\
     			\hline
     			
     			\makecell{6} & \multicolumn{2}{c|}{block ($16\times256\times6$)} \\
     			\hline
     			
     			\makecell{7} & \multicolumn{2}{c|}{block ($16\times128\times32$)} \\
     			\hline
     			
     			\makecell{8} & \multicolumn{2}{c|}{ConvTranspose ($16\times1\times64$)} \\
     			\hline
     			
     			\makecell{9} & \multicolumn{2}{c|}{FC ($16\times256$)} \\
     			\hline
     			
     			\makecell{10} & \multicolumn{2}{c|}{reshape ($16\times1\times256$)} \\
     			\hline
     			
     			\makecell{11} & \multicolumn{2}{c|}{FC ($16\times1\times256$)} \\
     			\hline
     			
     			\makecell{12} & \multicolumn{2}{c|}{$16\times1\times256$} \\
     			\hline
 			
 		\end{tabular}		
 	
 	\label{tab:archs_acwgangp_gen}
\end{table}

\begin{table}[H]
	\scriptsize
 	\caption{\textit{\textbf{Conditional GAN Architecture - Critic}}}
 	\centering
 		\begin{tabular}{| c | c | c | }
 		\hline
 			\thead{\textbf{Layer}} & \multicolumn{2}{c|}{\thead{\textbf{Critic}}} \\
 			\hline
 			\hline
 				\makecell{Input} & \multirow{3}{*}{$16\times1\times256$ (feature)}  & \makecell[cc]{16 (label)}\\
 				\cline{1-1}\cline{3-3}

     			\makecell{1} & & \makecell[cc]{embedding ($16\times256$)}\\
     			\cline{1-1}\cline{3-3}
     			
     			\makecell{2} &   & \makecell[cc]{reshape ($16\times1\times256$)}\\
     			\hline
     			
     			\makecell{3} & \multicolumn{2}{c|}{Concatenate ($16\times2\times256$)} \\
     			\hline
     			
     			\makecell{4} & \multicolumn{2}{c|}{Conv1d, ReLU ($16\times64\times128$)} \\
     			\hline
     			
     			\makecell{5} & \multicolumn{2}{c|}{block ($16\times128\times64$)} \\
     			\hline
     			
     			\makecell{6} & \multicolumn{2}{c|}{block ($16\times256\times32$)} \\
     			\hline
     			
     			\makecell{7} & \multicolumn{2}{c|}{block ($16\times512\times16$)} \\
     			\hline
     			
     			\makecell{7} & \multicolumn{2}{c|}{Conv1d ($16\times1\times7$)} \\
     			\hline
     			
     			\makecell{8} & \multicolumn{2}{c|}{FC ($16\times1\times7$)} \\
     			\hline
     			
     			\makecell{11} & \multicolumn{2}{c|}{FC ($16\times1\times1$)} \\
     			\hline
     			
     			\makecell{Output} & \multicolumn{2}{c|}{$16\times1\times1$} \\
     			\hline
 			
 		\end{tabular}		
 	
 	\label{tab:archs_acwgangp_critic}
\end{table}

\section{Experimental Description}

There are $15$ classes altogether in the MIT-BIH dataset. We picked $7$ classes, namely: \textbf{P} (Paced Beat), \textbf{A} (Atrial Premature Contraction),  \textbf{L} (Left Bundle Branch Block Beat), \textbf{N} (Normal Beat), \textbf{R} (Right Bundle Branch Block Beat), \textbf{f} (Fusion of Paced and Normal Beat), \textbf{j} (Nodal/Junctional Escape Beat). The support sets of the selected classes are shown in Table \ref{tab:mitbih_classes}. First, the data in each class are split randomly into the training and test sets by a split ratio of $0.9/0.1$. However, only $50\%$ of samples in each class are actually used in training because we wanted the classifier to train on a highly imbalanced dataset and perform purposely poorly due to an insufficient number of samples in minor classes. The number of samples \emph{actually} used in each set/class are also shown in Table \ref{tab:mitbih_classes}. The classifier used is the state-of-the-art (\emph{EcgResNet34}). It is trained on the imbalanced training set and the classification metrics were calculated. The metrics used are Precision/Recall/F1-score (per class, macro- and micro-averaged), total accuracy, confusion matrix, as well as Precision-Recall curve). This is our \emph{reference case} and its metric results were compared with the corresponding values for all the four augmented study cases. 

  \begin{table}[ht]
 	\caption{\textit{\textbf{Selected Classes From MIT-BIH Arrhythmia Dataset \\ Number of Samples}}}
 	\centering
 	\scriptsize
    
    \begin{minipage}{19mm}
      \begin{tabular}{| c | c |}
 		\hline
 		     \makecell [tc]{\thead{\textbf{Cl.}} \\ \phantom{'}} &        \makecell[tc]{\thead{\textbf{Total}} \\ \phantom{'}} \\
 			\hline
 			\hline
     		\makecell{P} & \makecell[tc]{7028} \\
     		\hline
     		\makecell{A} & \makecell[tc]{2546} \\ 
     		\hline
     		\makecell{L} & \makecell[c]{8075} \\ 
     		\hline
     		\makecell{N} &  \makecell[c]{75052} \\ 
     		\hline
     		\makecell{R} &  \makecell[c]{7259} \\ 
     		\hline
     		\makecell{f} &  \makecell[c]{982} \\ 
     		\hline
     		\makecell[tc]{j} &  \makecell[tc]{229} \\ 
     		\hline
 		\end{tabular}
    \end{minipage}
    \begin{minipage}{43mm}
      \begin{tabular}{| c | c |}
 		\hline
 			\makecell[tc]{\thead{\textbf{Samples Used}}\\ (train set)} & \makecell[tc]{\thead{\textbf{Samples Used}}\\ (test set)}\\
 			\hline
 			\hline
     			\makecell[tc]{3162} & \makecell[tc]{703}\\
     			\hline
     			\makecell[tc]{1145} & \makecell[tc]{255}\\ 
     			\hline
     			\makecell[tc]{3633} & \makecell[tc]{806}\\ 
     			\hline
     			\makecell[tc]{33773} & \makecell[tc]{7500}\\ 
     			\hline
     			\makecell[tc]{3266} & \makecell[tc]{726}\\ 
     			\hline
     			\makecell[tc]{441} & \makecell[tc]{99}\\ 
     			\hline
     			\makecell[tc]{103} & \makecell[tc]{23}\\ 
     			\hline
 		\end{tabular}
    \end{minipage}
    \label{tab:mitbih_classes}
  \end{table}

\paragraph{Templates}
For each class, a sample beat is selected visually by a domain expert as the template to represent the class. The template should have all the required patterns and morphological features which the domain expert determines.

\paragraph{Four Study Cases}
The generated beats by the two models, i.e., \textbf{\emph{(a)}} conditional and \textbf{\emph{(b)}} unconditional GANs, are used once as \textbf{\emph{(i)}} \emph{raw} or \emph{unscreened}, and once as \textbf{\emph{(ii)}} \emph{screened}. In the first scenario with \emph{raw} signals, all the generated beats are used whereas in \emph{screened} case, only the good quality beats are used, i.e., each generated beat is compared with the template of its class using the DTW distance function and discarded if its distance is greater than the threshold. Thus, four study cases were formed: \emph{\textbf{Case I}} - Conditional GAN, raw generated beats; \emph{\textbf{Case II}} - Conditional GAN, Screened generated beats; \emph{\textbf{Case III}} - Unconditional GAN, raw generated beats; \emph{\textbf{Case IV}} - Unconditional GAN, Screened generated beats.

\paragraph{Threshold and Distance Function}
Dynamic Time Warping (DTW) is used as the distance function. It takes two heartbeat $1D$-vectors as inputs and generates a scalar which represent the distance between the two inputs. A threshold is used in screening as the cutoff. The threshold was around $1.5$-$2$ for all classes except for class \emph{L} which was $5$. The reason for a higher threshold for class \emph{L} is that the generated data in this class were so noisy and screening with such a tight margin threshold were extremely time-consuming. 

\subsection{Platform}
A Dell Alienware with Intel i$9$-$9900$k at $3.6$ GHz ($8$ cores, $16$ threads) microprocessor, $64$ GB RAM, and NVIDIA GeForce RTX $2080$ Ti graphics card with $24$ GB RAM, and also a personal Dell G$7$ laptop with an Intel i$7$-$8750$H at $2.2$ GHz ($6$ cores, $12$ threads) microprocessor, $20$ GB of RAM, and NVIDIA GeForce $1060$ MaxQ graphics card with $6$ GB of RAM have been used in this study.
  
 The codes were written in Python $3.8$, and PyTorch $1.7.1$ was used as the main deep learning network library. The codes are available on the GitHub page of the paper (\url{https://github.com/mah533/Augmentation-of-ECG-Training-Dataset-with-CGAN})

\section{Results and Discussion}

\subsection{Samples of Generated Beats}
\begin{figure}[h]
 	\centering
 	\begin{tabular}{p{25mm} p{25mm} p{25mm}}
 		\makecell[cc]{\includegraphics[scale=0.56, trim=7mm 41.5mm 104.8mm 47.85mm,clip]{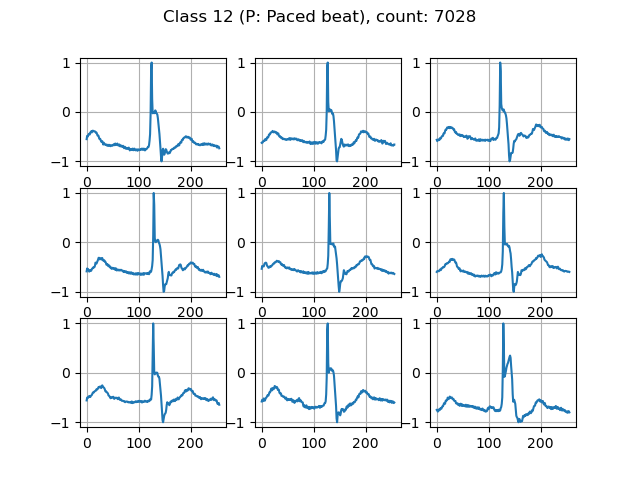}} & 
 		\makecell[cc]{\includegraphics[scale=0.56, trim=7mm 41.5mm 104.8mm 47.85mm,clip]{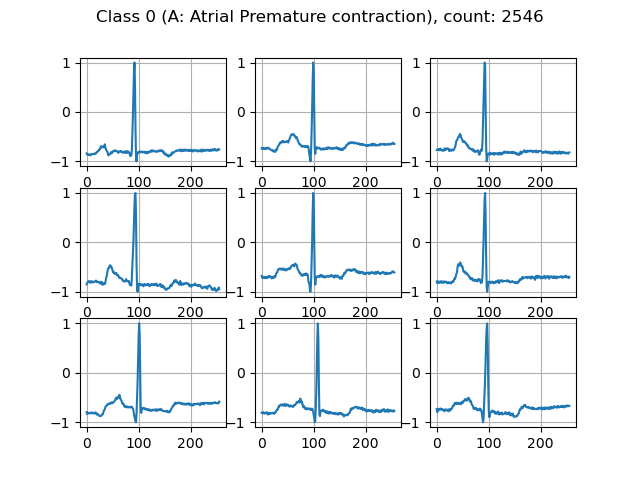}} &
 		\makecell[cc]{\includegraphics[scale=0.56, trim=7mm 41.5mm 104.8mm 47.85mm,clip]{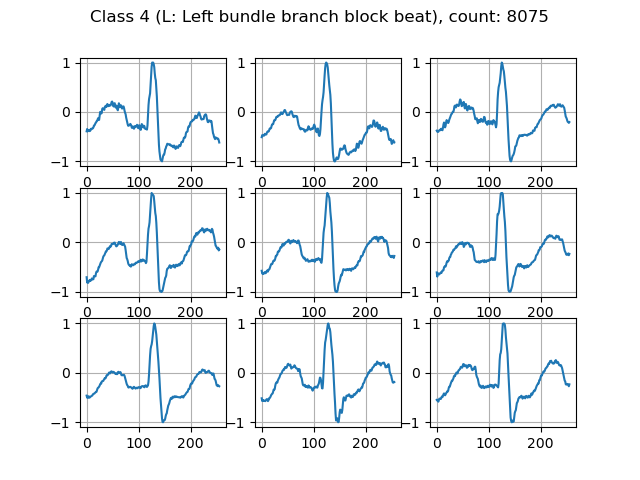}}
 		\\
 		\makecell[cc]{($a$) \scriptsize Class P} &
 		\makecell[cc]{($b$) \scriptsize Class A} &
 		\makecell[cc]{($b$) \scriptsize Class L}
        \\
 		\makecell[cc]{\includegraphics[scale=0.56, trim=7mm 41.5mm 104.8mm 47.6mm,clip]{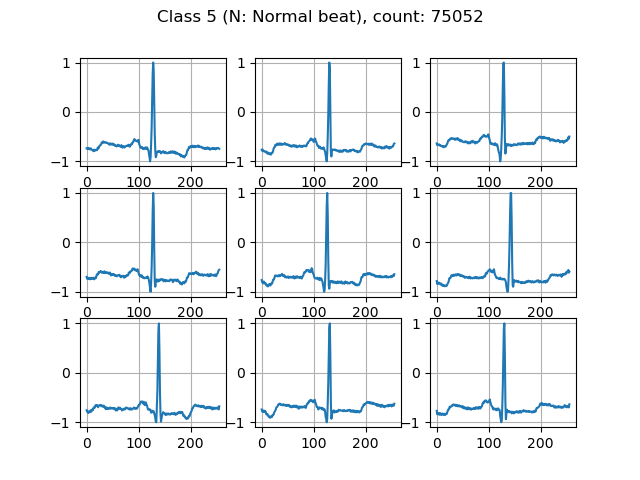}} & 
 		\makecell[cc]{\includegraphics[scale=0.56, trim=7mm 41.5mm 104.8mm 47.6mm,clip]{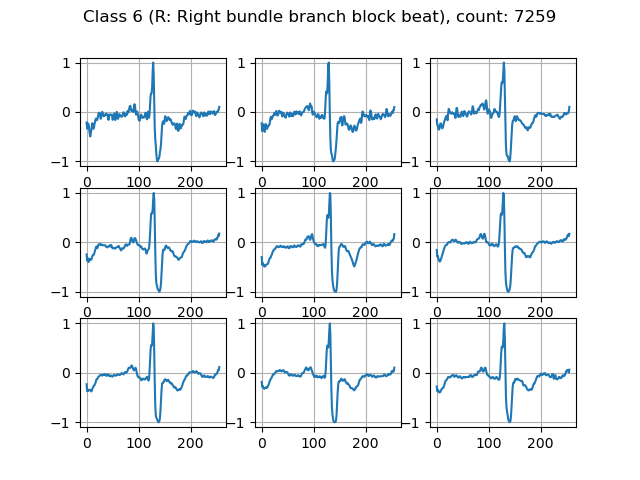}} &
 		\makecell[cc]{\includegraphics[scale=0.56, trim=7mm 41.5mm 104.8mm 47.6mm,clip]{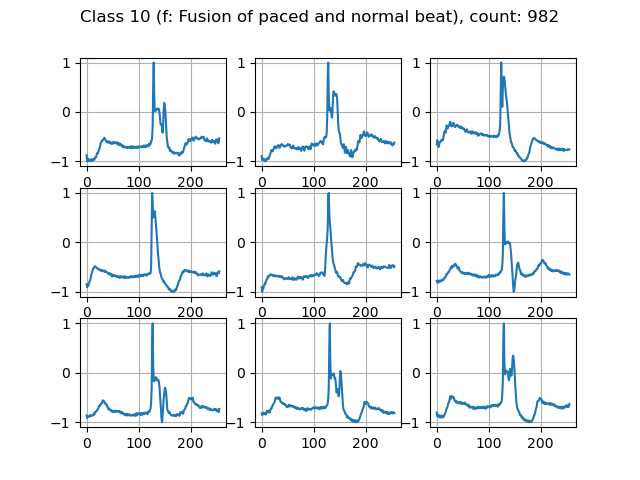}}
 		\\
 		\makecell[cc]{($d$) \scriptsize Class N} &
 		\makecell[cc]{($e$) \scriptsize Class R} &
 		\makecell[cc]{($f$) \scriptsize Class F}
 		\\
 		\makecell[cc]{\includegraphics[scale=0.56, trim=7mm 41.5mm 104.8mm 47.7mm,clip]{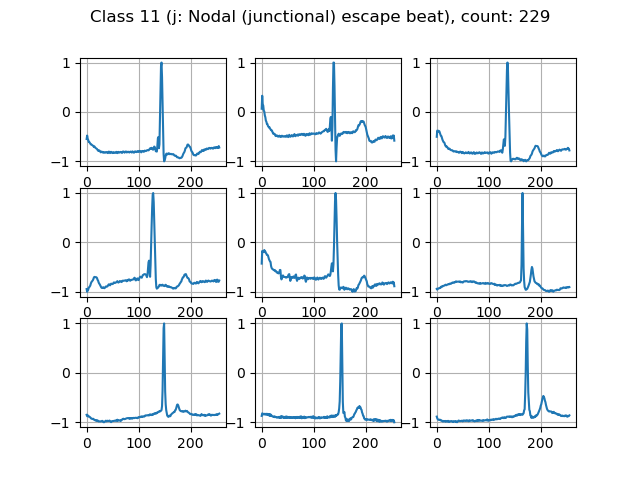}} & 	 & 
 		\\
 		\makecell[cc]{($g$) \scriptsize Class j} &  &
 	\end{tabular}
 	\caption{Samples of Generated Beats}
 	\label{fig:genbeats_samples}
 \end{figure}

A sample of the generated beats in each class is shown in Figure \ref{fig:genbeats_samples}. 

\subsection{Quality of Generated Beats Compared with Real Beats}

Table \ref{tab:gb_quality} summarizes the quality of the generated beats in the \emph{Cases I} through \emph{IV} plus that of real data, quantified by the DTW distance function and expressed as the average distance from the template in each class. Unlike screened cases (\emph{II} and \emph{IV}), Raw cases (\emph{I} and \emph{III}) show DTW values relatively higher than those of the real data almost in all classes, which is quite reasonable because in the screened cases, all the \emph{off} beats are discarded already. Evidently, the distribution of the set will change by the screening.

\begin{table}[h!]
	\scriptsize
 	\caption{\textit{\textbf{Quality of Generated Beats \& Real Data\\ Average DTW Distance from Template}}}
 	\centering
 		\begin{tabular}{| c | c | c | c | c | c |}
 		\hline
 	    	
 			\makecell[cc]{\textbf{Cl.}} & \makecell[cc]{\textbf{Case I} \\ (C.\footnote[1]{Conditional} Raw)} &  \makecell[cc]{\textbf{Case II}\\(C. Screened)} &  \makecell[cc]{\textbf{Case III} \\ (U.\footnote[2]{Unconditional} Raw) } &  \makecell[cc]{\textbf{Case IV} \\ (U. Screened)} & \makecell[cc]{\textbf{Real}\\ \textbf{Data}} \\
 			
 			\hline
 			\hline
 			
     			\makecell[tc]{P} & \makecell[tc]{5.99}  & \makecell[tc]{2.09} & \makecell[tc]{4.05} & \makecell[tc]{1.52} & \makecell[tc]{4.01}\\
     			\hline
     			
     			\makecell{A} & \makecell[tc]{6.56}  & \makecell[tc]{1.72} & \makecell[tc]{6.26} & \makecell[tc]{1.59} & \makecell[tc]{6.70}\\
     			\hline
     			
     			\makecell{L} & \makecell[cc]{5.00}  & \makecell[cc]{2.35} & \makecell[cc]{4.79} & \makecell[cc]{3.12} & \makecell[cc]{3.94}\\
     			\hline
     			
     			\makecell{R} & \makecell[cc]{3.37}  & \makecell[cc]{0.85} & \makecell[cc]{4.39} & \makecell[cc]{0.82} & \makecell[cc]{3.64}\\
     			\hline

     			\makecell{f} & \makecell[cc]{3.35}  & \makecell[cc]{2.14} & \makecell[cc]{2.47} & \makecell[cc]{2.00} & \makecell[cc]{2.80}\\
     			\hline
     			
     			\makecell{j} & \makecell[cc]{5.04}  & \makecell[cc]{2.14} & \makecell[cc]{3.47} & \makecell[cc]{2.09} & \makecell[cc]{3.01}\\
     			\hline
 		\end{tabular}		
 	
 	\label{tab:gb_quality}
\end{table}

\vspace{-5mm}

\begin{center}
\begin{table}[h!]
\scriptsize
\begin{tabular}{llll}
1: &  \hspace{-4mm} Conditional & \hspace{10mm} 2: & \hspace{-4mm} Unconditional
\end{tabular}
\end{table}
\end{center}

\subsection{Classification Results and Confusion Matrices}
\subsubsection{Reference Case}
\begin{figure}[H]
	\centering
	\includegraphics[scale=0.45, trim=5mm 50mm 10mm 23mm,clip]{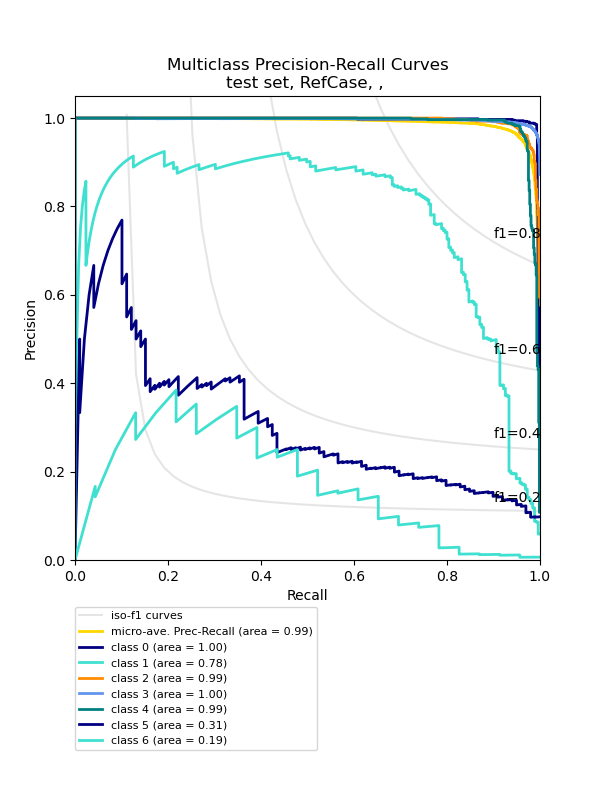}
	\caption{Precision-Recall Curves, Reference Case}
	\label{fig:Pr-Recall_ref}
\end{figure}

Figure  \ref{fig:Pr-Recall_ref} shows the multiclass Precision-Recall curve for the reference case (imbalanced dataset). The classification report and the confusion matrix for this case are shown in Tables \ref{tab:cl_rep_ref} and \ref{tab:cfmx_ref}, respectively. Since the training set was highly imbalanced, the poor performance was predictable, especially in the minority classes ($"f"$ and $"j"$). In class $"j"$ only $1$ sample out of $23$ $(4.35\%)$ is classified correctly and in class $"f"$, only $65.7\%$ of samples are classified correctly, whereas in other classes where the number of samples was high enough, more samples are correctly classified which is reflected in the confusion matrix diagonals.

\subsubsection{Augmented Training Sets (Cases \textbf{I} through \textbf{IV})}
\begin{figure}[H]
	\centering
	\includegraphics[scale=0.45, trim=5mm 10mm 10mm 24mm,clip] {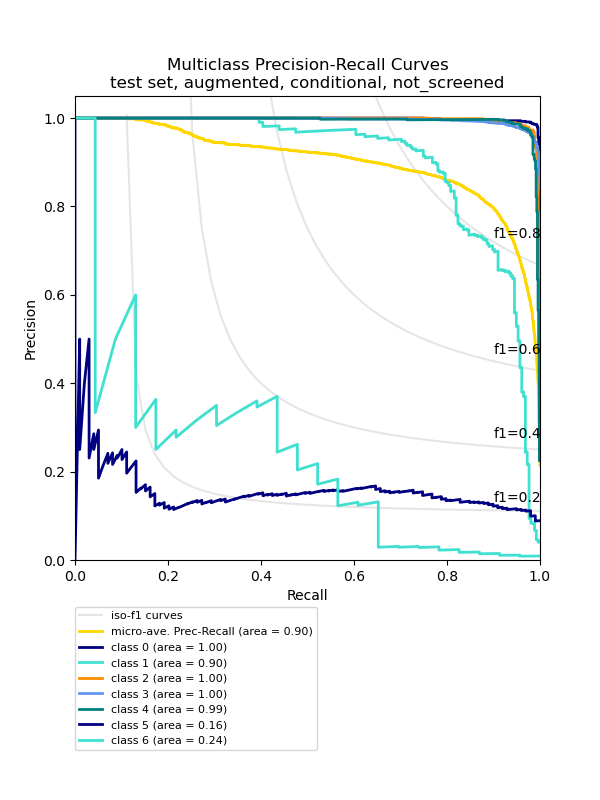}
	\caption{Precision-Recall Curves, Case \textit{\textbf{I}} (Conditional Raw)}
	\label{fig:Pr-Recall_cond_raw}
\end{figure}

\begin{figure}[H]
	\centering
	\includegraphics[scale=0.45, trim=5mm 50mm 10mm 24mm,clip] {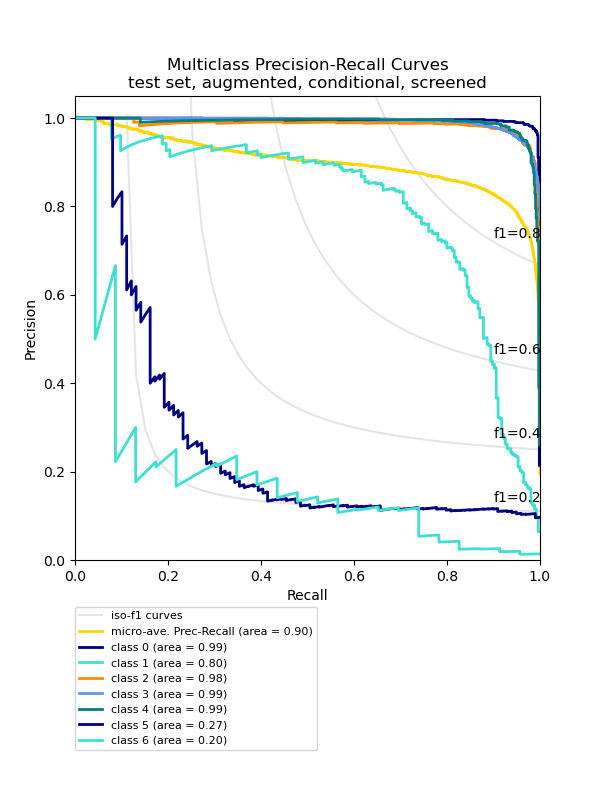}
	\caption{Precision-Recall Curves, Case \textit{\textbf{II}} (Conditional, Screened)}
	\label{fig:Pr-Recall_cond_screened}
\end{figure}

\begin{figure}[H]
	\centering
	\includegraphics[scale=0.45, trim=5mm 50mm 10mm 24mm,clip] {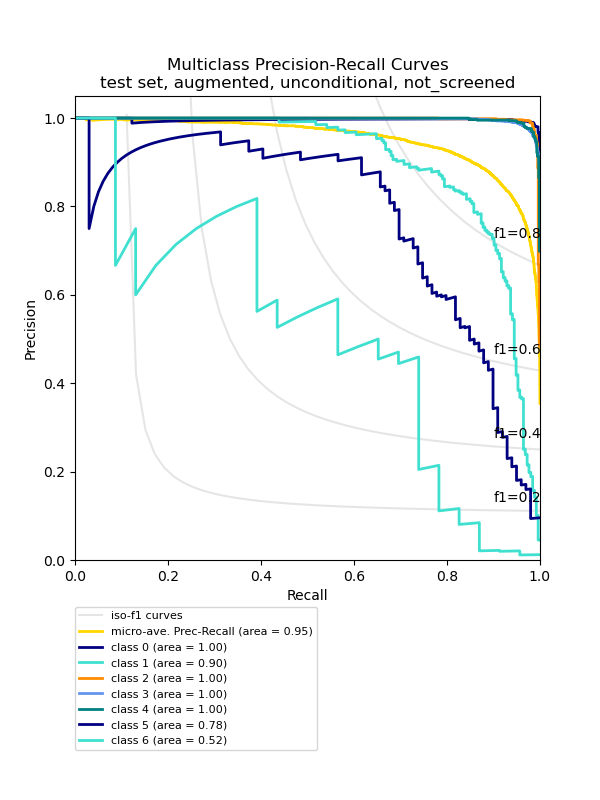}
	\caption{\textit{\textbf{Precision-Recall Curves, Case \textit{\textbf{III}} (Unconditional, Raw)}}}
	\label{fig:Pr-Recall_uncond_raw}
\end{figure}

\begin{figure}[H]
	\centering
	\includegraphics[scale=0.45, trim=5mm 50mm 10mm 24mm,clip] {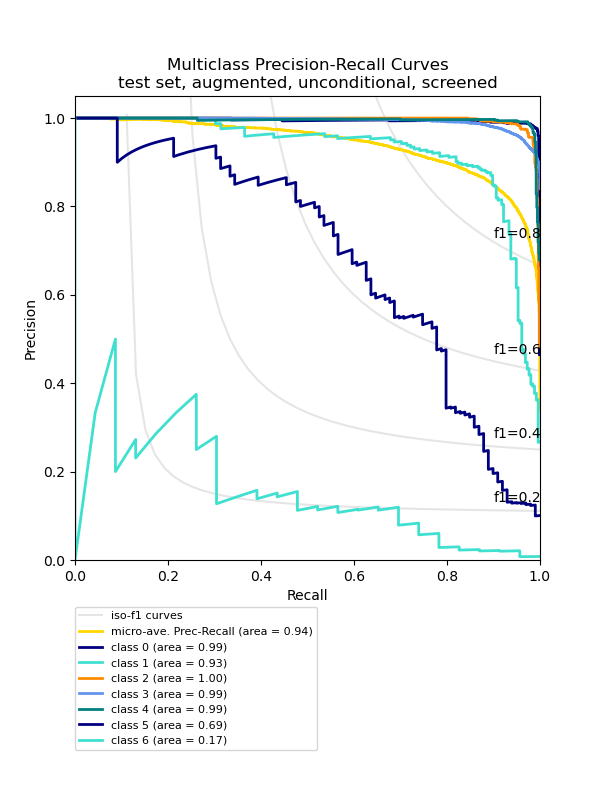}
	\caption{Precision-Recall Curves, Case \textit{\textbf{IV}} (Unconditional, Screened)}
	\label{fig:Pr-Recall_uncond_screened}
\end{figure}

The original imbalanced dataset is augmented with the generated synthetic beats in cases \textit{\textbf{I}} through \textit{\textbf{IV}} and each is balanced up to a total of $10,000$ samples per class. Then the classifier \textit{EcgResNet34} is trained on the four augmented datasets. The same unseen test set is used to evaluate the improvements achieved by augmentation in classification performances compared to the \emph{Reference Case}. The classification metrics and the confusion matrix results for the four study cases are shown in Tables \ref{tab:cl_rep_cond_raw} to \ref{tab:cfmx_case_iV}. Also the corresponding multiclass Precision-Recall curves are shown in Figures \ref{fig:Pr-Recall_cond_raw} through \ref{fig:Pr-Recall_uncond_screened}. 

The results show a uniform and significant improvement in all metrics in all cases. The same improvement is seen in the corresponding confusion matrix of all the four cases. The number of samples which are classified correctly, especially in minor classes, improved significantly.

\begin{table}[H]
	\scriptsize
 	\caption{\textit{\textbf{Classification Report \\ Reference Case (Imbalanced Dataset)}}}
 	\centering
 		\begin{tabular}{| c | c | c | c | c |}
 		\hline
 			\thead{\textbf{Cl.}} & \thead{\textbf{Precision}} &  \thead{\textbf{Recall}} & \makecell[tc]{\thead{\textbf{F1-Score}}} & \makecell[tc]{\thead{\textbf{Support}}}\\
 			\hline
 			\hline
 			
     			\makecell{P} & \makecell[tc]{0.95}  & \makecell[tc]{0.99} & \makecell[tc]{0.97} & \makecell[tc]{703}\\
     			\hline
     			
     			\makecell{A} & \makecell[tc]{0.90}  & \makecell[tc]{0.66} & \makecell[tc]{0.76} & \makecell[tc]{255}\\
     			\hline
     			
     			\makecell{L} & \makecell[tc]{0.93}  & \makecell[tc]{0.92} & \makecell[tc]{0.93} & \makecell[tc]{806}\\
     			\hline
     			
     			\makecell{N} & \makecell[tc]{0.96}  & \makecell[tc]{0.99} & \makecell[tc]{0.98} & \makecell[tc]{7500}\\
     			\hline
     			
     			\makecell{R} & \makecell[tc]{0.99}  & \makecell[tc]{0.91} & \makecell[tc]{0.95} & \makecell[tc]{726}\\
     			\hline
     			
     			\makecell{f} & \makecell[tc]{0.82}  & \makecell[tc]{0.09} & \makecell[tc]{0.16} & \makecell[tc]{99}\\
     			\hline
     			
     			\makecell{j} & \makecell[tc]{0.00}  & \makecell[tc]{0.00} & \makecell[tc]{0.00} & \makecell[tc]{23}\\
     			\hline
     			\hline
     			
     			\makecell{Accuracy} & \makecell[tc]{}  & \makecell[tc]{} & \makecell[tc]{0.96} & \makecell[tc]{10112}\\
     			\hline
     			
     			\makecell{Macro avg} & \makecell[tc]{0.79}  & \makecell[tc]{0.65} & \makecell[tc]{0.68} & \makecell[tc]{10112}\\
     			\hline
     			
     			\makecell{Micro avg} & \makecell[tc]{0.96}  & \makecell[tc]{0.96} & \makecell[tc]{0.96} & \makecell[tc]{10112}\\
     			\hline
 		\end{tabular}
 	\label{tab:cl_rep_ref}
\end{table}

\begin{table}[H]
	\scriptsize
 	\caption{\textit{\textbf{Confusion Matrix (\%) \\ Reference Case (Imbalanced Dataset)}}}
 	\centering
 		\begin{tabular}{| c | c | c | c | c | c | c | c |}
 		\hline
 			\textbf{} & \textbf{P} &  \textbf{A} & \textbf{L} & \textbf{N} & \textbf{R} & \textbf{f} & \textbf{j}\\
 			\hline
 			\hline
     			\makecell[tc]{\textbf{P}} & \makecell[tc]{98.2}  & \makecell[tc]{0.14} & \makecell[tc]{0.85} & \makecell[tc]{0.43}  & \makecell[tc]{0.00} & \makecell[tc]{0.43} & \makecell[tc]{0.00}\\
     			\hline
     			
     			\makecell{\textbf{A}} & \makecell[tc]{0.00}  & \makecell[tc]{67.1} & \makecell[tc]{3.53} & \makecell[tc]{27.1}  & \makecell[tc]{2.35} & \makecell[tc]{0.00} & \makecell[tc]{0.00}\\
     			\hline
     			
     			\makecell{\textbf{L}} & \makecell[cc]{0.00}  & \makecell[cc]{0.00} & \makecell[cc]{99.6} & \makecell[cc]{0.25}  & \makecell[cc]{0.00} & \makecell[cc]{0.12} & \makecell[cc]{0.00}\\
     			\hline
     			
     			\makecell{\textbf{N}} & \makecell[cc]{0.03}  & \makecell[cc]{0.07} & \makecell[cc]{0.73} & \makecell[cc]{98.8}  & \makecell[cc]{0.31} & \makecell[cc]{0.03} & \makecell[cc]{0.00}\\
     			\hline
     			
     			\makecell{\textbf{R}} & \makecell[cc]{0.00}  & \makecell[cc]{0.14} & \makecell[cc]{0.00} & \makecell[cc]{2.35}  & \makecell[cc]{97.5} & \makecell[cc]{0.00} & \makecell[cc]{0.00}\\
     			\hline
     			
     			\makecell{\textbf{f}} & \makecell[cc]{5.05} & \makecell[cc]{0.00}  & \makecell[cc]{7.07} & \makecell[cc]{20.2} & \makecell[cc]{2.02}  & \makecell[cc]{65.7}  & \makecell[cc]{0.00}\\
     			\hline
     			
     			\makecell{\textbf{j}}& \makecell[cc]{0.00}  & \makecell[cc]{0.00} & \makecell[cc] {0.00}& \makecell[cc]{73.9}  & \makecell[cc]{21.7} & \makecell[cc]{0.00} & \makecell[cc]{4.35}\\
     			\hline
 		\end{tabular}
 	\label{tab:cfmx_ref}
\end{table}


\begin{table}[H] 
	\scriptsize
 	\caption{\textit{\textbf{Classification Report \\ Case I: Conditional GAN, Raw Gen. Beats}}}
 	\centering
 		\begin{tabular}{| c | c | c | c | c |}
 		\hline
 			\thead{\textbf{Cl.}} & \thead{\textbf{Precision}} &  \thead{\textbf{Recall}} & \makecell[cc]{\thead{\textbf{F1-Score}}} & \makecell[cc]{\thead{\textbf{Support}}}\\
 			\hline
 			\hline
     			\makecell{P} & \makecell[cc]{0.95}  & \makecell[cc]{1.00} & \makecell[cc]{0.98} & \makecell[cc]{703}\\
     			\hline
     			\makecell{A} & \makecell[cc]{0.62}  & \makecell[cc]{0.92} & \makecell[cc]{0.74} & \makecell[cc]{255}\\
     			\hline
     			\makecell{L} & \makecell[cc]{0.86}  & \makecell[cc]{0.99} & \makecell[cc]{0.92} & \makecell[cc]{806}\\
     			\hline
     			\makecell{N} & \makecell[cc]{1.00}  & \makecell[cc]{0.90} & \makecell[cc]{0.95} & \makecell[cc]{7500}\\
     			\hline
     			\makecell{R} & \makecell[cc]{0.76}  & \makecell[cc]{0.99} & \makecell[cc]{0.86} & \makecell[cc]{726}\\
     			\hline
     			\makecell{f} & \makecell[cc]{0.35}  & \makecell[cc]{0.89} & \makecell[cc]{0.50} & \makecell[cc]{99}\\
     			\hline
     			\makecell{j} & \makecell[cc]{0.12}  & \makecell[cc]{0.74} & \makecell[cc]{0.20} & \makecell[cc]{23}\\
     			\hline
     			\hline
     			\makecell{Accuracy} & \makecell[cc]{}  & \makecell[cc]{} & \makecell[cc]{0.92} & \makecell[cc]{10112}\\
     			\hline
     			\makecell{Macro avg} & \makecell[cc]{0.67}  & \makecell[cc]{0.92} & \makecell[cc]{0.74} & \makecell[cc]{10112}\\
     			\hline
     			\makecell{Micro avg} & \makecell[cc]{0.95}  & \makecell[cc]{0.92} & \makecell[cc]{0.93} & \makecell[cc]{10112}\\
     			\hline
 		\end{tabular}
 	\label{tab:cl_rep_cond_raw}
\end{table}

\begin{table}[H]
	\scriptsize
 	\caption{\textit{\textbf{Confusion Matrix (\%) \\ Case I: Conditional GAN, Raw Gen. Beats}}}
 	\centering
 		\begin{tabular}{| c | c | c | c | c | c | c | c |}
 		\hline
 			\textbf{} & \textbf{P} &  \textbf{A} & \textbf{L} & \textbf{N} & \textbf{R} & \textbf{f} & \textbf{j}\\
 			\hline
 			\hline
     			\makecell[tc]{\textbf{P}} & \makecell[tc]{99.9}  & \makecell[tc]{0.14} & \makecell[tc]{0.00} & \makecell[tc]{0.00}  & \makecell[tc]{0.00} & \makecell[tc]{0.00} & \makecell[tc]{0.00}\\
     			\hline
     			
     			\makecell{\textbf{A}} & \makecell[tc]{0.78}  & \makecell[tc]{92.2} & \makecell[tc]{0.39} & \makecell[tc]{1.18}  & \makecell[tc]{3.92} & \makecell[tc]{0.39} & \makecell[tc]{1.18}\\
     			\hline
     			
     			\makecell{\textbf{L}} & \makecell[cc]{0.12}  & \makecell[cc]{0.87} & \makecell[cc]{98.89} & \makecell[cc]{0.00}  & \makecell[cc]{0.00} & \makecell[cc]{0.00} & \makecell[cc]{0.12}\\
     			\hline
     			
     			\makecell{\textbf{N}} & \makecell[cc]{0.24}  & \makecell[cc]{1.79} & \makecell[cc]{1.65} & \makecell[cc]{89.75}  & \makecell[cc]{2.85} & \makecell[cc]{2.07} & \makecell[cc]{1.65}\\
     			\hline
     			
     			\makecell{\textbf{R}} & \makecell[cc]{0.14}  & \makecell[cc]{0.14} & \makecell[cc]{0.14} & \makecell[cc]{0.14}  & \makecell[cc]{99.3} & \makecell[cc]{0.14} & \makecell[cc]{0.00}\\
     			\hline
     			
     			\makecell{\textbf{f}} & \makecell[cc]{11.2} & \makecell[cc]{0.00}  & \makecell[cc]{0.00} & \makecell[cc]{0.00} & \makecell[cc]{0.00}  & \makecell[cc]{88.8}  & \makecell[cc]{0.00}\\
     			\hline
     			
     			\makecell{\textbf{j}}& \makecell[cc]{0.00}  & \makecell[cc]{0.00} & \makecell[cc] {0.00}& \makecell[cc]{8.70}  & \makecell[cc]{0.00} & \makecell[cc]{17.4} & \makecell[cc]{73.9}\\
     			\hline
 		\end{tabular}		
 	
 	\label{tab:cfmx_case_i}
\end{table}


\begin{table}[H]
	\scriptsize
 	\caption{\textit{\textbf{Classification Report \\ Case II: Conditional GAN, Screened Gen. Beats}}}
 	\centering
 		\begin{tabular}{| c | c | c | c | c |}
 		\hline
 			\thead{\textbf{Cl.}} & \thead{\textbf{Precision}} &  \thead{\textbf{Recall}} & \makecell[cc]{\thead{\textbf{F1-Score}}} & \makecell[cc]{\thead{\textbf{Support}}}\\
 			\hline
 			\hline
     			\makecell{P} & \makecell[cc]{0.99}  & \makecell[cc]{0.95} & \makecell[cc]{0.97} & \makecell[cc]{703}\\
     			\hline
     			
     			\makecell{A} & \makecell[cc]{0.52}  & \makecell[cc]{0.96} & \makecell[cc]{0.67} & \makecell[cc]{255}\\
     			\hline
     			
     			\makecell{L} & \makecell[cc]{0.91}  & \makecell[cc]{0.99} & \makecell[cc]{0.95} & \makecell[cc]{806}\\
     			\hline
     			
     			\makecell{N} & \makecell[cc]{1.00}  & \makecell[cc]{0.92} & \makecell[cc]{0.96} & \makecell[cc]{7500}\\
     			\hline
     			
     			\makecell{R} & \makecell[cc]{0.86}  & \makecell[cc]{0.99} & \makecell[cc]{0.92} & \makecell[cc]{726}\\
     			\hline
     			
     			\makecell{f} & \makecell[cc]{0.43}  & \makecell[cc]{0.95} & \makecell[cc]{0.59} & \makecell[cc]{99}\\
     			\hline
     			
     			\makecell{j} & \makecell[cc]{0.22}  & \makecell[cc]{0.78} & \makecell[cc]{0.34} & \makecell[cc]{23}\\
     			\hline
     			\hline
     			
     			\makecell{Accuracy} & \makecell[cc]{}  & \makecell[cc]{} & \makecell[cc]{0.94} & \makecell[cc]{10112}\\
     			\hline
     			
     			\makecell{Macro avg} & \makecell[cc]{0.70}  & \makecell[cc]{0.93} & \makecell[cc]{0.77} & \makecell[cc]{10112}\\
     			\hline
     			
     			\makecell{Micro avg} & \makecell[cc]{0.96}  & \makecell[cc]{0.94} & \makecell[cc]{0.94} & \makecell[cc]{10112}\\
     			\hline
 		\end{tabular}
 	\label{tab:cl_rep_cond_screened}
\end{table}

\begin{table}[H]
	\scriptsize
 	\caption{\textit{\textbf{Confusion Matrix (\%) \\ Case II: Conditional GAN, Screened Gen. Beats}}}
 	\centering
 			\begin{tabular}{| c | c | c | c | c | c | c | c |}
 		\hline
 			\textbf{} & \textbf{P} &  \textbf{A} & \textbf{L} & \textbf{N} & \textbf{R} & \textbf{f} & \textbf{j}\\
 			\hline
 			\hline
     			\makecell{\textbf{P}} & \makecell[cc]{95.3}  & \makecell[cc]{1.14} & \makecell[cc]{0.14} & \makecell[cc]{0.14}  & \makecell[cc]{0.00} & \makecell[cc]{2.99} & \makecell[cc]{0.28}\\
     			\hline
     			
     			\makecell{\textbf{A}} & \makecell[cc]{0.00}  & \makecell[cc]{96.1} & \makecell[cc]{0.39} & \makecell[cc]{2.75}  & \makecell[cc]{0.00} & \makecell[cc]{0.78} & \makecell[cc]{0.00}\\
     			\hline
     			
     			\makecell{\textbf{L}} & \makecell[cc]{0.00}  & \makecell[cc]{0.37} & \makecell[cc]{98.5} & \makecell[cc]{0.74}  & \makecell[cc]{0.12} & \makecell[cc]{0.25} & \makecell[cc]{0.00}\\
     			\hline
     			
     			\makecell{\textbf{N}} & \makecell[cc]{0.05}  & \makecell[cc]{2.88} & \makecell[cc]{1.00} & \makecell[cc]{92.4}  & \makecell[cc]{1.55} & \makecell[cc]{1.27} & \makecell[cc]{0.80}\\
     			\hline
     			
         			\makecell{\textbf{R}} & \makecell[cc]{0.00}  & \makecell[cc]{0.41} & \makecell[cc]{0.00} & \makecell[cc]{0.41}  & \makecell[cc]{98.9} & \makecell[cc]{0.00} & \makecell[cc]{0.28}\\
     			\hline
     			
     			\makecell{\textbf{f}} & \makecell[cc]{5.05}  & \makecell[cc]{0.00} & \makecell[cc]{0.00} & \makecell[cc]{0.00}  & \makecell[cc]{0.00} & \makecell[cc]{95.0} & \makecell[cc]{0.00}\\
     			\hline
     			
     			\makecell{\textbf{j}} & \makecell[cc]{0.00}  & \makecell[cc]{0.00} & \makecell[cc]{0.00} & \makecell[cc]{8.70}  & \makecell[cc]{0.00} & \makecell[cc]{13.0} & \makecell[cc]{78.3}\\
     			\hline
 		\end{tabular}
 	\label{tab:cfmx_case_II}
\end{table}


\begin{table}[H]
	\scriptsize
 	\caption{\textit{\textbf{Classification Report \\ Case III: Unconditional GAN, Raw Gen. Beats}}}
 
 	\centering
 		\begin{tabular}{| c | c | c | c | c |}
 		\hline
 			\thead{\textbf{Cl.}} & \thead{\textbf{Precision}} &  \thead{\textbf{Recall}} & \makecell[cc]{\thead{\textbf{F1-Score}}} & \makecell[cc]{\thead{\textbf{Support}}}\\
 			\hline
 			\hline
     				\makecell{P} & \makecell[cc]{0.98}  & \makecell[cc]{0.99} & \makecell[cc]{0.99} & \makecell[cc]{703}\\
     			\hline
     			
     			\makecell{A} & \makecell[cc]{0.48}  & \makecell[cc]{0.98} & \makecell[cc]{0.64} & \makecell[cc]{255}\\
     			\hline
     			
     			\makecell{L} & \makecell[cc]{0.94}  & \makecell[cc]{0.99} & \makecell[cc]{0.96} & \makecell[cc]{806}\\
     			\hline
     			
     			\makecell{N} & \makecell[cc]{1.00}  & \makecell[cc]{0.92} & \makecell[cc]{0.96} & \makecell[cc]{7500}\\
     			\hline
     			
     			\makecell{R} & \makecell[cc]{0.90}  & \makecell[cc]{1.00} & \makecell[cc]{0.94} & \makecell[cc]{726}\\
     			\hline
       			
     			\makecell{f} & \makecell[cc]{0.40}  & \makecell[cc]{0.92} & \makecell[cc]{0.56} & \makecell[cc]{99}\\
     			\hline
     			
     			\makecell{j} & \makecell[cc]{0.15}  & \makecell[cc]{0.87} & \makecell[cc]{0.25} & \makecell[cc]{23}\\
     			\hline
     			\hline
     			
     			\makecell{Accuracy} & \makecell[cc]{}  & \makecell[cc]{} & \makecell[cc]{0.93} & \makecell[cc]{10112}\\
     			\hline
     			
     			\makecell{Macro avg} & \makecell[cc]{0.69}  & \makecell[cc]{0.95} & \makecell[cc]{0.76} & \makecell[cc]{10112}\\
     			\hline
     			
     			\makecell{\textbf{\textit{Micro avg}}} & \makecell[cc]{0.97}  & \makecell[cc]{0.93} & \makecell[cc]{0.94} & \makecell[cc]{10112}\\
     			\hline
 		\end{tabular}
 	\label{tab:cl_rep_Uncon_raw}
\end{table}

\begin{table}[H]
	\scriptsize
 	\caption{\textit{\textbf{Confusion Matrix (\%) \\ Case III: Unconditional GAN, Raw Gen. Beats}}}
 	\centering
 			\begin{tabular}{| c | c | c | c | c | c | c | c |}
 		\hline
 			\textbf{} & \textbf{P} &  \textbf{A} & \textbf{L} & \textbf{N} & \textbf{R} & \textbf{f} & \textbf{j}\\
 			\hline
 			\hline
     			\makecell{\textbf{P}} & \makecell[cc]{99.0}  & \makecell[cc]{0.28} & \makecell[cc]{0.00} & \makecell[cc]{0.14}  & \makecell[cc]{0.00} & \makecell[cc]{0.57} & \makecell[cc]{0.00}\\
     			\hline
     			
     			\makecell{\textbf{A}} & \makecell[cc]{0.00}  & \makecell[cc]{97.65} & \makecell[cc]{0.39} & \makecell[cc]{0.39}  & \makecell[cc]{0.78} & \makecell[cc]{0.39} & \makecell[cc]{0.39}\\
     			\hline
     			
     			\makecell{\textbf{L}} & \makecell[cc]{0.00}  & \makecell[cc]{0.37} & \makecell[cc]{98.9} & \makecell[cc]{0.12}  & \makecell[cc]{0.12} & \makecell[cc]{0.50} & \makecell[cc]{0.00}\\
     			\hline
     			
     			\makecell{\textbf{N}} & \makecell[cc]{0.07}  & \makecell[cc]{3.51} & \makecell[cc]{0.69} & \makecell[cc]{91.5}  & \makecell[cc]{1.05} & \makecell[cc]{1.65} & \makecell[cc]{1.51}\\
     			\hline
     			
     			\makecell{\textbf{R}} & \makecell[cc]{0.00}  & \makecell[cc]{0.14} & \makecell[cc]{0.00} & \makecell[cc]{0.00}  & \makecell[cc]{99.6} & \makecell[cc]{0.00} & \makecell[cc]{0.28}\\
     			\hline
     			
     			\makecell{\textbf{f}} & \makecell[cc]{6.06}  & \makecell[cc]{0.00} & \makecell[cc]{1.01} & \makecell[cc]{0.00}  & \makecell[cc]{1.01} & \makecell[cc]{91.9} & \makecell[cc]{0.00}\\
     			\hline
     			
     			\makecell{\textbf{j}} & \makecell[cc]{0.00}  & \makecell[cc]{0.00} & \makecell[cc]{0.00} & \makecell[cc]{4.35}  & \makecell[cc]{4.35} & \makecell[cc]{4.35} & \makecell[cc]{87.0}\\
     			\hline
 		\end{tabular}
 	\label{tab:cfmx_case_III}
\end{table}


\begin{table} [H]
	\scriptsize
 	\caption{\textit{\textbf{Classification Report \\ Case IV: Unconditional GAN, Screened}}}
 	\centering
 		\begin{tabular}{| c | c | c | c | c |}
 		\hline
 			\thead{\textbf{Cl.}} & \thead{\textbf{Precision}} &  \thead{\textbf{Recall}} & \makecell[cc]{\thead{\textbf{F1-Score}}} & \makecell[cc]{\thead{\textbf{Support}}}\\
 			\hline
 			\hline
     			\makecell{P} & \makecell[cc]{0.93}  & \makecell[cc]{1.00} & \makecell[cc]{0.96} & \makecell[cc]{703}\\
     			\hline
     			
     			\makecell{A} & \makecell[cc]{0.58}  & \makecell[cc]{0.96} & \makecell[cc]{0.72} & \makecell[cc]{255}\\
     			\hline
     			
     			\makecell{L} & \makecell[cc]{0.91}  & \makecell[cc]{1.00} & \makecell[cc]{0.95} & \makecell[cc]{806}\\
     			\hline
     			
     			\makecell{N} & \makecell[cc]{1.00}  & \makecell[cc]{0.93} & \makecell[cc]{0.97} & \makecell[cc]{7500}\\
     			\hline
     			
     			\makecell{R} & \makecell[cc]{0.92}  & \makecell[cc]{0.99} & \makecell[cc]{0.96} & \makecell[cc]{726}\\
     			\hline
     			
     			\makecell{f} & \makecell[cc]{0.48}  & \makecell[cc]{0.84} & \makecell[cc]{0.61} & \makecell[cc]{99}\\
     			\hline
     			
     			\makecell{j} & \makecell[cc]{0.25}  & \makecell[cc]{0.87} & \makecell[cc]{0.39} & \makecell[cc]{23}\\
     			\hline
     			\hline
     			
     			\makecell{Accuracy} & \makecell[cc]{}  & \makecell[cc]{} & \makecell[cc]{0.95} & \makecell[cc]{10112}\\
     			\hline
     			
     			\makecell{Macro avg} & \makecell[cc]{0.72}  & \makecell[cc]{0.94} & \makecell[cc]{0.79} & \makecell[cc]{10112}\\
     			\hline
     			
     			\makecell{Micro avg} & \makecell[cc]{0.96}  & \makecell[cc]{0.95} & \makecell[cc]{0.95} & \makecell[cc]{10112}\\
     			\hline
 		\end{tabular}
 	\label{tab:cl_rep_Uncon_screened}
\end{table}

\begin{table}[H]
	\scriptsize
 	\caption{\textit{\textbf{Confusion Matrix (\%) \\ Case IV: Unconditional GAN, Screened}}}
 	\centering
 			\begin{tabular}{| c | c | c | c | c | c | c | c |}
 		\hline
 			\textbf{} & \textbf{P} &  \textbf{A} & \textbf{L} & \textbf{N} & \textbf{R} & \textbf{f} & \textbf{j}\\
 			\hline
 			\hline
     			\makecell{\textbf{P}} & \makecell[tl]{99.9}  & \makecell[tl]{0.14} & \makecell[tl]{0.00} & \makecell[tl]{0.00}  & \makecell[tl]{0.00} & \makecell[tl]{0.00} & \makecell[tl]{0.00}\\
     			\hline
     			
     			\makecell{\textbf{A}} & \makecell[tl]{0.39}  & \makecell[tl]{96.1} & \makecell[tl]{0.00} & \makecell[tl]{1.96}  & \makecell[tl]{1.57} & \makecell[tl]{0.00} & \makecell[tl]{0.00}\\
     			\hline
     			
     			\makecell{\textbf{L}} & \makecell[tl]{0.12}  & \makecell[tl]{0.25} & \makecell[tl]{99.5} & \makecell[tl]{0.00}  & \makecell[tl]{0.12} & \makecell[tl]{0.00} & \makecell[tl]{0.00}\\
     			\hline
     			
     			\makecell{\textbf{N}} & \makecell[tl]{0.56}  & \makecell[tl]{2.36} & \makecell[tl]{0.99} & \makecell[tl]{93.4}  & \makecell[tl]{0.72} & \makecell[tl]{1.20} & \makecell[tl]{0.73}\\
     			\hline
     			
         			\makecell{\textbf{R}} & \makecell[tl]{0.00}  & \makecell[tl]{0.14} & \makecell[tl]{0.00} & \makecell[tl]{0.14}  & \makecell[tl]{99.2} & \makecell[tl]{0.00} & \makecell[tl]{0.55}\\
     			\hline
     			
     			\makecell{\textbf{f}} & \makecell[tl]{12.1}  & \makecell[tl]{0.00} & \makecell[tl]{1.01} & \makecell[tl]{2.02}  & \makecell[tl]{0.00} & \makecell[tl]{83.8} & \makecell[tl]{1.01}\\
     			\hline
     			
     			\makecell{\textbf{j}} & \makecell[tl]{0.00}  & \makecell[tl]{0.00} & \makecell[tl]{0.00} & \makecell[tl]{8.70}  & \makecell[tl]{0.00} & \makecell[tl]{4.35} & \makecell[tl]{87.0}\\
     			\hline
 		\end{tabular}
 	\label{tab:cfmx_case_iV}
\end{table}

\subsection{Net Improvements in True Positives}
Another metric that is used in this study is the \textit{total} net improvements in true positives (main diagonals in the confusion matrix). For calculation of this metric \textit{all} the elements on the main diagonal of the confusion matrix in the reference case are subtracted from the corresponding elements in each case and the results are summed up algebraically. For the improvements in \emph{minor classes} the same method has been used, but only minor classes ("\textit{i}" and "\textit{j}") are considered. We believe this simple metric, which focuses only on true positives, is more tangible for the purpose of this study, because here we are only interested in those synthetic samples which can be classified correctly. As seen from Table \ref{tab:comp_scores}, \textbf{\emph{Case III}} produces the most improvements in true positives (\emph{total} and \emph{minor-classes}), which is consistent with the other two metrics.

\begin{table} [H]
	\scriptsize
 	\caption{\textit{\textbf{Net Improvement in True Positives (\%)}}}
 	\centering
 			\begin{tabular}{| c | c | c | c | c |}
 		\hline
 			\textbf{-} & \textbf{Case I} &  \textbf{Case II} & \textbf{Case III} & \textbf{Case IV} \\
 			\hline
 			\hline
     			\makecell{\textbf{Total}}  & \makecell[tl]{111.4}  & \makecell[tc]{123.3} & \makecell[tc]{\textbf{\emph{134.3}}} & \makecell[tc]{127.6} \\
     			\hline
     			
     			\makecell{\textbf{Minor-Classes}} & \makecell[tc]{92.68}  & \makecell[tc]{103.2} & \makecell[tc]{\textbf{\emph{108.9}}} & \makecell[tc]{100.8}\\
     			\hline
     			
 		\end{tabular}		
 	
 	\label{tab:comp_scores}
\end{table}

\subsection{Precision and Recall }
In this study we are firstly interested in samples which can be classified correctly, i.e., maximizing number of true positives (TP). Secondly, we like to minimize number of false positives (FP), i.e., those samples which are mistakenly classified as correct. Thus, \begin{small}$\frac{FP}{TP}$\end{small} is to be minimized. Therefore, Precision score \begin{small}$(Pr=\frac{TP}{TP+FP}=\frac{1}{1+FP/TP})$\end{small} relates to the purpose of this study more than Recall score, as in the abundance of the generated data, it does not really matter if some of the \emph{good} samples are overlooked (false negatives). However, Precision-Recall curve captures the trade-off between the two scores at different thresholds. Micro-averaged Precision scores favor \textbf{\emph{Case III}} with the highest value of $0.97$ (Table \ref{tab:cl_rep_Uncon_raw}). Also, Precision-Recall curve selects \textbf{\emph{Case III}} with highest area under the curve (Figure \ref{fig:Pr-Recall_uncond_raw}). Both scores are consistent with the Net Improvement score. 

\section{Conclusion}
In this paper, we developed a conditional AC-WGAN-GP model \emph{in one dimensional form} for the first time and implemented it along with a non-conditional WGAN-GP model to investigate the impact of data augmentation in arrhythmia detection. We employed the two models to generate synthetic heartbeats samples in $7$ arrhythmia classes from the MIT-BIH Arrhythmia dataset. Two scenarios have been considered for each model: \textbf{\emph{(i)}} raw data, i.e., all the generated data have been used for augmentation and \textbf{\emph{(ii)}} screened data, i.e., only good-quality generated beats (determined by the DTW distance function) have been used. Thus, four study cases are developed. The state-of-the-art classifier (\textit{EcgResNet34}) is employed to investigate the effectiveness of data augmentation. First, the classifier is trained on the original imbalanced dataset (reference case) and then on the four augmented datasets. Then, the trained classifier in each case is used to classify the same holdout unseen data (test set) and then the classification metrics are compared. The metrics used are \textbf{\emph{(i)}} micro-averaged Precision score, \textbf{\emph{(ii)}} multiclass Precision-Recall graph and \textbf{\emph{(iii)}} the net improvements in the number of true positives (\emph{total} and \emph{minor-class}). We believe that this last metric is the most suitable one for the purpose of this study, because we are interested only in the number of synthetic beats that can be classified correctly. All the three metrics consistently select \textbf{\emph{Case III}}, i.e., unconditional GAN with no screening.

It might seem that screened cases should produce better improvements in true positives as the low-quality beats are already discarded from training set. However, during screening and by discarding some of the generated beats, the distribution changes. Since GAN models implicitly mimic the distribution of the training set in the generated data, the distribution of the generated data would be different from the original real dataset in screened cases. Thus, the augmented set is composed of two sets of data with two different distributions. This would negatively impact the training, optimization and convergence of the parameters of the classifier model. 

\newpage


\vspace{12pt}
\color{red}

\end{document}